\documentstyle[preprint,aps,epsfig]{revtex}

\newcommand{\be}{\begin{equation}}
\newcommand{\ee}{\end{equation}}
\newcommand{\beq}{\begin{eqnarray}}
\newcommand{\eeq}{\end{eqnarray}}

\begin{document}

\title{Noise Delays Bifurcation in a Positively Coupled Neural Circuit}
\author{Boris Gutkin $^{1}$, Tim Hely $^{2}$, Juergen Jost $^{2,3}$}

\address{ 1.  Unite de Neurosciences Integratives et Computationalles, 
INAF, CNRS, Avenue de la Terrasse, 91198, Gif-sur-Yvette,
Cedex, France. Email: Boris.Gutkin@iaf.cnrs-gif.fr 2. Santa Fe Institute, 1399 Hyde Park Road, Santa Fe, NM,
87501. 3. Max Planck Institute for Mathematics in the Sciences,
Inselstr. 22-26, D-04103 Leipzig, Germany.}
\maketitle

\begin{abstract}
  { We report a noise induced delay of bifurcation in a simple
    pulse-coupled neural circuit.  We study the behavior of two neural
    oscillators, each individually governed by saddle-node dynamics,
    with reciprocal excitatory synaptic connections.  In the
    deterministic circuit, the synaptic current amplitude acts as a
    control parameter to move the circuit from a mono-stable regime
    through a bifurcation into a bistable regime. In this regime
    stable sustained anti-phase oscillations in both neurons coexist
    with a stable rest state. We introduce a small amount of random
    current into both neurons to model possible randomly arriving
    synaptic inputs. We find that such random noise delays the onset
    of bistability, even though in decoupled neurons noise tends to
    advance bifurcations and the circuit has only excitatory
    coupling. We show that the delay is dependent on the level of
    noise and suggest that a curious stochastic ``anti-resonance'' is
    present.  \vskip 15pt \noindent PACS numbers:
    87.10.+e,87.18.Bb,87.18.Sn,87.19.La }
\end{abstract}

\section{Introduction}
The effects of random currents on the firing behavior of real and
model neurons have received a considerable amount of attention in
neurobiology and physics literature
\cite{Segundo94,Mainen95,CollinsGrigg97,Chialvo97,Longtin97,Lee98,Rodriguez98,Mar99}.
Several experimental results indicate that in vivo neural spike trains
seem to be excessively noisy, with interspike interval distribution
showing 1/f spectra \cite{Usher95}.  However, other \textit{in vitro}
experiments have shown that noisy stimuli can produce highly reliable
firing with the neuron locking onto large range variations of the
noise \cite{Mainen95,Reich97}. A number of theoretical studies have
attempted to reconcile such seemingly disparate results by studying
the dynamics of neural networks with additive noise, showing that high
variance firing behavior can arise in networks of threshold elements
\cite{vanVreeswijk98}. At the same time additive noise in oscillating
networks of more realistic neurons destabilizes synchronous and phase
locked behavior, producing complicated spatiotemporal patterns
\cite{GolombAmitai97}. These simulation results have appeared in the
context of a body of literature that has delved into the effects of
noise on the response of excitable and oscillatory non-linear
dynamical systems. In particular a number of investigators have
considered what happens to single neurons and circuits of neurons when
noise perturbs periodically modulated input signals. Experimental work
has identified noise induced signal amplification and resonance in a
number of preparations e.g. \cite{CollinsGrigg96}.  Theoretical
analyses have successfully explained such findings employing the
language of stochastic resonance developed originally for general
multi-stable dynamical systems. There the enhancement of the
subthreshold stimuli and encoding of stimulus structure had a
non-linear relationship with the noise amplitude, resulting in a
signal-to-noise ratio relationship with a pronounced peak. Noise
effects have also been studied in the context of indigenous
oscillations in neural models, focusing on the so-called ``autonomous
stochastic resonance'' \cite{Longtin97}.  For example, a recent report
by Lee shows noise induced coherence resonance in a Hodgkin-Huxley
model, with pre-cursors of the sub-critical Hopf bifurcation revealed
by the action of random currents \cite{Lee98}. In this sense noise
``advanced'' the bifurcation. Similar effects have also been found in
a generic saddle-node driven oscillator where noise advances the onset
of oscillations and upregulates the mean frequency
\cite{Rappel94,Pikovsky97}

Although pulse coupled or synaptically coupled neural networks have
received much recent attention with regard to their dynamics
\cite{Maass97b,Lin98} and computational power \cite{Maass97a}, we
believe that this Letter is the first attempt to look in detail at the
effects of noise on the onset of synaptically sustained firing in such
networks. That is, circuits of intrinsically quiescent neurons where
activity occurs purely due to the recurrent synaptic interactions.  To
our knowledge, almost all efforts to study the interplay of noise and
neural oscillators report noise induced \textbf{increase} in firing
and advancement of bifurcations, e.g. \cite{Rappel94}. In this light
our finding is rather intriguing since we observe a noise induced
\textbf{delay} of bifurcation in a purely positively coupled circuit
of neural oscillators. We also observe a phenomenon that may be termed
``stochastic anti-resonance'', since the delay of bifurcation depends
non-linearly on the noise level.  Our analysis of this system leads us
to conclude that the relative width of the attractor basins for the
quiescent and persistent firing states is the key factor in
determining whether stochastic resonance has a delaying, neutral, or
advancing effect on the bifurcation.

Below we summarize the dynamics of the spiking neuron used in this
circuit (the $\theta$-neuron), and analyze the case of two coupled
cells in the regimes of weak and strong excitatory coupling in both
noise free and noisy simulations.  Since we believe that the
phenomenon we observe is generic for circuits of recurrently coupled
spiking neurons, first we describe the stochastic anti-resonance
phenomenon observed in this simple circuit.

Figure \ref{noise-bif}A, upper trace shows the firing patterns of two
cells whose spiking behavior results from mutual excitatory
synapses. The cells are initially quiescent (they are not
intrinsically spiking) and their activity results from an initial
external input to one cell.  Activity can be terminated by small
levels of noise (Figure \ref{noise-bif}A, middle trace), whilst
increased noise levels cause intermittent firing (Figure
\ref{noise-bif}A, lower trace). Figure \ref{noise-bif}B plots the
probability ($M_{1}$) of observing firing in the last 200 msecs of a
2000 msec run over an ensemble of 1000 sample paths. The x-axis plots
the strength of the synaptic coupling ($g_{s}$). In the noise free
circuit, $g_{s}^{*}$ is the critical value of coupling above which
sustained firing occurs (i.e. $M_{1}$=0 for $g_{s}<g_{s}^*$, $M_{1}$=1
for $g_{s}\geq g_{s}^{*}$). Since the synaptically sustained firing
apears with a non-zero frequency, we suspect that the bifurcation here
is of a sub-critical Hopf type.  At small noise levels (Figure
\ref{noise-bif}B, traces 1,2), increasing the noise amplitude
progressively shifts the curves of $M_{1}$ to the right with respect
to the noise-free case.  This behavior is surprising as addition of
small amounts of noise for a single autonomously spiking
$\theta$-neuron induces the opposite effect - noise advanced
bifurcation (see \cite{Gutkin98}). The effect has been described in a
generic saddle-node oscillator in \cite{Rappel94}.  Above a critical
noise value, the onset of sustained firing is advanced back to the
left (Figure \ref{noise-bif}B, traces 3 and higher).  Thus the
bifurcation is delayed for low noise amplitudes and advanced with
higher noise.  Figure \ref{noise-bif}{C} shows that there is a
non-linear relationship between the amount of injected noise and the
firing probability. Here we plot the value of $g_{s}=g_{s}^{2/3}$ at
which continuous sustained firing is observed in 2/3 of the sample
paths, the same points are marked on the probability plots in Figure
\ref{noise-bif}B.  Adding small amounts of noise moves the probability
curves to the right.  This can be viewed as a probabilistic signature
of a delay of the bifurcation. As the noise amplitude grows, the
bifurcation is delayed further, and the test point $g_{s}^{2/3}$ occurs
at higher $g_{s}$ values. As the noise is increased further, noise
fluctuations are strong enough to induce intermittent firing.  Both
the probability curves and the location of the test point then move
back to the left towards the noise-free value. If we consider the
sustained firing as signal (perturbed by noise), this resembles
stochastic resonance, however here the net effect of noise is
``negative''.

It should be noted that this effect is not restricted to the dynamics
of the $\theta$-neuron.  All aspects of noise induced delay of
bifurcation seen above also occur in a circuit where each of the cells
is modeled with a more complicated conductance based, Hodgkin-Huxley
model for a pyramidal neuron \cite{Traub96} (simulations
not shown).  This is not surprising, since this model can be readily
reduced to the $\theta$-neuron which we now describe.

\section{The $\theta$-neuron}
The $\theta$-neuron model developed by Ermentrout and Gutkin
\cite{Ermentrout96,Gutkin98} is derived from the observation that wide 
class of neuronal models of cortical neurons,
 based on the electrophysiological model of Hodgkin and
Huxley show a saddle-node type bifurcation at a critical parameter
value. This parameter determines the dynamical behavior of the
solutions of the corresponding system of ordinary differential
equations. General dynamical systems theory tells us that the
qualitative behavior in some neighborhood of the bifurcation point
(which may be quite large as it extends up to the next bifurcation or
other dynamic transition) is governed by the reduction of the system
to the center manifold. In the present case of the saddle-node
bifurcation which is the simplest bifurcation type, this leads to the
following differential equation \begin{equation} \frac{dx}{dt}=\lambda
  + x^2.
\end{equation} Here, the bifurcation parameter $\lambda$ is considered
as the input to the neuron while $x$ records its activity. Obviously,
a solution to this equation tends to infinity in finite time. This is
considered as a spiking event, and the initial values are then reset
to $-\infty$. In order to have a model that does not exhibit such
formal singularities, one introduces a phase variable $\theta$ that is
$2\pi$-periodic via \begin{equation} x=\tan(\frac{\theta}{2}).
\end{equation} $\theta$ is then a variable with domain
the unit circle $S^1$, and a spike now corresponds to a period of
$\theta$. Spikes are no longer represented  by transitions
through infinity, but by changes of some discrete topological
invariant. 
The original differential equation is then transformed into
\begin{equation} \frac{d\theta}{dt}=(1-\cos(\theta)) +
  (1+\cos(\theta))\lambda.
\end{equation} Due to the nonlinearity of the transformation from $x$
to $\theta$, the input $\lambda$ is no longer additive. In fact, it is easy to show 
that $(1+cos\theta)$ is the 
phase resetting function for the model \cite{Ermentrout96}. As before,
the bifurcation occurs at $\lambda=0$. There, we have precisely one
rest point, namely $\theta=0$ which is degenerate.  In any case, the
sensitivity to the input $\lambda$ is highest at $\theta=0$ and lowest
at $\theta=\pi$ which according to the derivation of our equation is
considered as the spike point. When $\lambda$ is positive, the
equation does not have any rest point. In this case, $\theta$
continues to increase all the time, and the neuron is perpetually
firing. When $\lambda$ is negative, however, there are two rest
points, a stable one denoted by $\theta_r$ and an unstable one
$\theta_t>\theta_r$. If $\theta$ is larger than
$\theta_t$ it increases until it completes a period and comes to rest
at $\theta_r+2\pi$ which is identified with $\theta_r$ as we are
working on the unit circle $S^1$. Thus, if the phase is above the
threshold value $\theta_t$, a spike occurs and the neuron
returns to rest. So far, we have tacitly assumed that the input
$\lambda$ is constant. We now consider the situation where the input
can be decomposed as \begin{equation} \lambda=\beta+\sigma\eta,
\end{equation} where $\beta$ is a constant term, the so-called bias,
while $\eta$ is (white) noise and $\sigma$ its intensity. In this
case, sufficiently strong noise can occasionally push the phase $\theta$ beyond the
threshold value $\theta_t$ causing intermittent firing (Figure
\ref{noise-bif}C). Equation 3 now becomes a canonical stochastic saddle-node 
oscillator which has been studied in 
Rappel \& Wooten and Gutkin \& Ermentrout \cite{Gutkin98}. 

\section{Coupled neurons} 
\begin{figure}[htb]
  \begin{center}
    \leavevmode 
    \epsfxsize=150mm
    \epsfbox{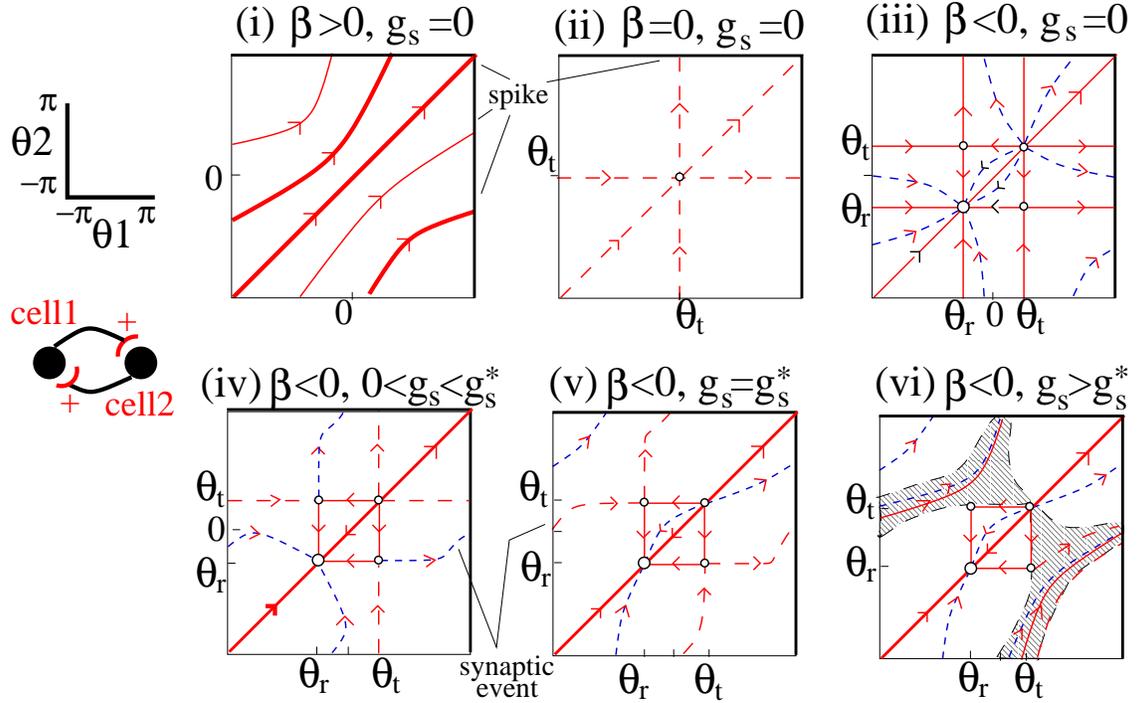}
    \caption{Different states of the network for
various values of the intrinsic excitability of the cells, $\beta$,
and the coupling strength, $g_{s}$. Axes plot the phase
$(\theta_{1},\theta_{2})$ of each cell.  See text for full details.}
  \label{thetaStates}
  \end{center}
\end{figure}

We now consider the situation where we have
two neurons (distinguished by subscripts $i=1,2$). The
dynamics then takes place on the product of two circles, i.e. on a
two-dimensional torus $T$, represented by the square
$[-\pi,\pi]\times[-\pi,\pi]$ in the plane, with periodic boundary
identifications. We first consider the simple case of two uncoupled,
noise-free neurons ($\sigma_{1}=\sigma_{2}=0$) with the same bias
$\beta$. Their dynamics are independent.  In the phase plot shown in
Figure \ref{thetaStates}(i) the diagonal is always an invariant curve,
corresponding to synchronized activity of the two neurons. If
$\beta>0$, both neurons continue to fire, although their phase
difference, if not 0 initially, is not constant, due to the
nonlinearity of the differential equation governing it. If $\beta=0$,
$(0,0)$ is a degenerate rest point (Figure \ref{thetaStates}(ii)). The
two curves $\theta_1=\theta_2=0$ are homoclinic orbits and all flow
lines eventually terminate at this fixed point.  One or both neurons
will spike before returning to rest if their initial phase is between
0 and $\pi$.

If $\beta<0$ (Figure \ref{thetaStates}(iii)), we have four fixed
points - the attractor $(\theta_1=\theta_2=\theta_r)$, the repeller
$(\theta_1=\theta_2=\theta_t)$, and the two saddles where one of the
neurons has its phase at $\theta_r$ (rest) and the other one at
$\theta_t$ (threshold). Some special heteroclinic orbits are given by
the straight lines where one of the two neurons stays at $\theta_t$
while the other one moves from the threshold to the rest value,
spiking if its initial phase was above threshold. All other flow lines
terminate at the attractor.  We now add an interaction term $s_i g_s$
to the input of neuron $i$. $s_i$ is considered as the synaptic input
from neuron $j$ to neuron $i$ ($i\neq j$) and $g_s$ is the synaptic
intensity. (One could also study the case of a single neuron $i$ for
which $s_i$ represents synaptic self-coupling, but here we are
interested in the case of two coupled neurons).  A precise equation
for $s_i$ can be derived from electrophysiological models, however for
our qualitative study we only need the characteristic features that it
stays bounded between 0 and 1.  Typically, it is peaked near the spike
of neuron $j$, i.e.  where $\theta_j=\pi$. With this interaction term,
the equation for neuron $i$ then becomes \begin{equation}
\frac{d\theta_i}{dt}=(1-\cos(\theta_i))+(1+\cos(\theta_i))(\beta + g_s
s_i + \sigma\eta). \end{equation} Since $s_i$ represents the input
that neuron $i$ receives from neuron $j$, $s_i$ should essentially be
considered as a function of the phase $\theta_j$ of $j$. Once more, we
first consider the situation without noise, i.e.  $\sigma=0$ (although
our final aim is to understand the effect of noise on the dynamic
behavior of the coupled neurons). We also assume that we are in the
excitable region, i.e. $\beta<0$. $g_s$ is assumed to be positive
(excitatory coupling), and so the coupling counteracts the effect of
the bias to a certain extent, a crucial difference being, however,
that the synaptic input to each neuron is time dependent, in contrast
to the constant bias. If $g_s$ is sufficiently small, the qualitative
situation does not change compared to the case without coupling,
i.e. $g_s=0$. We still have a heteroclinic orbit from the saddle
$(\theta_1=\theta_t, \theta_2=\theta_r)$ to the attractor
$(\theta_r,\theta_r)$, although $\theta_2$ does not stay constant
anymore along that orbit, but increases first a little due to the
input from neuron 1 before it descends again to the rest value.
(Figure \ref{thetaStates}(iv)). (Of course, we also get such an orbit
with the roles of the two neurons reversed; in fact, the dynamical
picture is always invariant under reflection across the diagonal,
i.e.under exchanging the two neurons.) If $g_s$ reaches some critical
value $g_s^*$, however, the heteroclinic orbit starting at
$(\theta_t,\theta_r)$ does not terminate anymore at the attractor, and
the value of the phase of neuron 2 is increased so much by the
synaptic interaction that it reaches the other saddle
$(\theta_r,\theta_t)$ (Figure \ref{thetaStates}{v}). Besides two
heteroclinic orbits that go from the repeller to the two saddles as
before, all other orbits still terminate at the attractor
$(\theta_r,\theta_r)$, for $g_s=g_s^{*}$. If $g_s$ is increased beyond
$g_s^{*}$, however, the heteroclinic orbit between the two saddles
mutates into a stable attractor (Figure \ref{thetaStates}(vi)). It
corresponds to sustained asynchronous firing of the two neurons. In
fact, if the phase difference between the two neurons is too small,
the dynamics converges towards the double rest point (except in some
region in the vicinity of the node), and both neurons stop
firing. This is caused by the fact that when the two neurons are close
to synchrony, neither cell is sensitive enough to its synaptic input
to maintain firing (an effective refractory period).  Conversely, if
they are out of synchrony, a single spike can induce the second neuron
to fire at a time when the first one is close to rest, and sensitive
to synaptic input itself. If $g_s$ is only slightly above the critical
value, the basin of attraction of that limit cycle will still be
relatively small, but as $g_s$ is increased further, the basin grows
in size until eventually it is larger than the basin of attraction of
the double rest point. On the basis of the preceding analysis, it is
now straightforward to predict the effect of noise. If $g_s$ is only
slightly above the critical value $g_s^{*}$, a small amount of noise
is more likely to kick the dynamics out of the narrow basin of
attraction of the asynchronous limit cycle and into the large basin of
the double rest point than vice versa. In effect, a small noise level
increases the critical parameter value required for the qualitative
transition to sustained asynchronous firing. A larger amount of noise,
however, has the potential to move the dynamics from the rest point
into the basin of attraction of the asynchronous limit cycle. Once in
that basin, the neurons will fire.  Thus, for large noise in that
regime, one will observe that the neurons will fire, perhaps with some
intermissions spent near the double rest point. So, a larger value of
noise will cause intermittent periods of sustained firing of the two
neurons even at somewhat smaller values of $g_s$. In effect, it
decreases the value of the critical parameter.  Thus, we observe a
genuinely nonlinear effect of the noise level $\sigma$ (Figure
\ref{noise-bif}E). For values of the coupling $g_s$ that are
substantially larger than the critical value $g_s^{*}$, even small
amounts of noise have a good chance of perturbing the dynamics out of
the attracting vicinity of the double rest point into the attracting
region of the asynchronous limit cycle. This will further enhance the
sustained asynchronous firing pattern of the two neurons.

\section{Conclusions}

In this work we report a new and unusual effect of noise in a simple
neural circuit. When the sustained oscillations in the circuit are
induced by recurrent excitatory coupling, small noise levels can exert
a strong influence on the circuit dynamics, often abolishing the
firing.  The probability of observing sustained firing has been used
to characterize the transition from quiescent to oscillatory behavior.
Figure \ref{noise-bif}B clearly shows that in this system, noise
delays this transition.  Noise induced delay of bifurcation can
therefore occur in a completely positively coupled circuit. The same
noise has the exact opposite effect of advancing the bifurcation when
it is applied to a single autonomously firing neuron.  The paradoxical
effect of noise in this circuit can be understood by considering the
structure of its phase plane - and in particular the width of the
attractor basins for the sustained antiphase oscillations.  When the
width of the attractor basin is small, small levels of noise can
perturb the system into the larger basin of the stable quiescent
state. However, transitions in the opposite direction from the
rest-state to a sustained firing state can only occur when noise
fluctuations reach a critical value.  Above this value, transitions
into the firing state begin to counteract transitions into the
quiescent state.  Alternatively, as the coupling strength increases,
the basin of attraction for the sustained firing solution grows at the
expense of the quiescent state.  The negative (bifurcation-delaying)
effect of the noise is then eliminated.  In this system low levels of
noise effectively act as a switch to turn off otherwise continuous
firing behavior.  Alternatively, low levels of noise ensure that
sustained firing can only take place above a critical coupling
threshold. In this way, small amounts of noise may in fact help to
reduce overall noise levels by eliminating the formation of spurious
attractors.  It has yet to be determined whether this effective
noise-induced control mechanism can be observed in large ensembles of
coupled neurons.

Funding was provided by National Science Foundation Bioinformatics
Postdoctoral Fellowship (B.S.G.) and the Santa Fe Institute (T.H. and J.J.)
The authors thank Cosma Shalizi for helpful discussions.

\newpage

\begin{figure}[htb]
  \begin{center}
    \leavevmode 
    \epsfxsize=150mm
    \epsfbox{fig1-ndb-bw.epsf}
    \caption{Asynchronous synaptically sustained oscillations in a
    positively coupled 2-cell circuit. A. Upper trace: the sustained
    firing in the noise free circuit. Middle trace: the sustained
    firing can be terminated by the action of small amplitude
    noise. Lower trace: larger amplitude noise induces an intermittent
    firing pattern. Here the noise injected into the two neurons is
    completely correlated, but the results are qualitatively identical
    for uncorrelated noises. B. Increasing noise delays sustained
    firing for low noise levels (traces 1,2) and advances firing for
    higher noise levels (traces 3,4,5), the horizontal dashed line and
    the numbers mark the test points $g_{s}^{2/3}$. C. Addition
    of noise has a non-linear effect on sustained firing in this
    coupled circuit. Here we plot the location of the test points
    $g^{2/3}_s$ See text for further details.}
  \label{noise-bif}
  \end{center}
\end{figure}

\newpage

\end{document}